\documentclass{article}
\usepackage{amsmath}
\usepackage{latexsym}
\usepackage{epsfig}
\usepackage{graphicx}

\newcommand{\clR}{\mathcal{R}}

\newcommand{\clP}{{\cal P}}

\newcommand{\clW}{{\cal W}}
\newcommand{\clL}{{\cal L}}
\newcommand{\clF}{{\cal F}}
\newcommand{\clT}{{\cal T}}
\newcommand{\clC}{{\cal C}}

\newcommand{\clE}{\mathcal{E}}

\newcommand{\prt}{\partial}

\newcommand{\frD}{D_{\rm fr}}

\newcommand{\bfx}{\mathbf{x}}
\newcommand{\bfk}{\mathbf{k}}
\newcommand{\rgl}{\rangle}
\newcommand{\lgl}{\langle}

\newcommand{\vep}{\varepsilon}
\newcommand{\be}{\begin{equation}}
\newcommand{\ee}{\end{equation}}
\newcommand{\bea}{\begin{eqnarray}}
\newcommand{\eea}{\end{eqnarray}}

\begin{document}



\title{CONTINUOUS TIME RANDOM WALK AND MIGRATION PROLIFERATION
DICHOTOMY OF BRAIN CANCER}

\author{A. IOMIN \\
Department of Physics,  Technion, Haifa 32000, Israel\\
\\
Published in BRL: DOI: 10.1142/S1793048014500052}


\maketitle


\begin{abstract}
A theory of fractional kinetics of glial cancer cells is
presented. A role of the migration-proliferation dichotomy in the
fractional cancer cell dynamics in the outer-invasive zone is
discussed an explained in the framework of a continuous time
random walk. The main suggested model is based on a construction
of a 3D comb model, where the migration-proliferation dichotomy
becomes naturally apparent and the outer-invasive zone of glioma
cancer is considered as a fractal composite with a fractal
dimension $\frD<3$.

\end{abstract}

\section{Introduction}

Brain tumors result from the uncontrolled growth of abnormal
cells, destruction of normal tissues, and invasion of vital
organs. These processes can be subdivided into many types based on
several classification characteristics and involve any of the cell
types found in the brain, such as neurons, glial cells,
astrocytes, or cells of the meninges \cite{MCB,AANS}. The
mechanisms behind cancer progression result from the accumulation
of one or a few specific mutations that disrupt biological
pathways like growth factor signaling, DNA damage repair, cell
cycle, apoptosis and cellular adhesion \cite{hanahan}. Among all
possible cancer cell genotypes, leading to six main alternations
of malignant growth \cite{hanahan}, cell motility and invasion are
the most important for the present consideration.

Glioma is one of the most recalcitrant brain disease, with an
optimal therapy treatment survival period of 15 month and most
tumors recur within 9 months of initial treatment
\cite{stupp1,stupp2}. One of the main possible reason of such
devastating manifestation is the migration proliferation dichotomy
of cancer cells. This phenomenon has been firstly observed at
clinical investigations \cite{Giese1,Giese2}, where it has been
shown that in the outer invasive zone glioma cancer cells
possesses a property of high motility, while the proliferation
rate of these migratory cells is essentially lower than in the
tumor core. This anti-correlation between proliferation and
migration of cancer cells, also known as the "Go or Grow"
hypothesis (see discussions in \cite{Corcoran,garay}), suggests
that cell division and cell migration are temporally exclusive
phenotypes \cite{Giese1}. The phenomenon that tumor cells defer
proliferation for cell migration was also experimentally
demonstrated\footnote{This kind of migration-proliferation
dichotomy was also found at metastatic behavior of breast cancer
\cite{breastcancer}.} in \cite{garay18,garay26,garay37}. The
switching process between these two phenotypes is still not well
understood. Moreover, it should be mentioned that conflicting data
appear in the literature concerning the Go or Grow hypothesis;
details of discussions on this can be found in
\cite{Corcoran,garay}.


Extensive theoretical modelling follow this finding. A switching
process between these two phenotypes still is not well understood,
and many efforts are directed to develop relevant models with
relevant mechanisms of switching of the glioma cells, resulting in
several phenomenological models. Comprehensive discussions of
these models one can find in
\cite{Khain,Deutsch,Chauviere,kolobov,fir2011}. It was suggested
by Khain \textit{et al} \cite{Khain,khain1} that the motility of
cancer cells is a function of their density. Multi-parametric
modelling of the phenotype switching was considered in
\cite{athale}. The agent-based approach to simulate multi-scale
glioma growth and invasion was used in \cite{Zhang1,Zhang2}.
Subdiffusive cancer development on a comb was studied in
\cite{iom2006}, where a continuous time random walk (CTRW) was
firstly suggested for metastatic cancer development. A stochastic
approach for the proliferation-migration switching involving only
two parameters was proposed in \cite{fi07,fi08} where the
transport of cancer cells was formulated in terms of the CTRW, as
well. `Go or Grow' mechanism was proposed in \cite{Deutsch}, where
the transition to invasive tumor phenotypes can be explained on
the basis of the oxygen shortage in the environment of a growing
tumor. Phenotypic switching due to density effect was also
suggested in \cite{Chauviere,Ch2}. Both numerical and analytical
approaches were developed in \cite{kolobov} to study the glioma
propagation in the framework of reaction-diffusion equations,
where the phenotype switching depends on oxygen in a threshold
manner. Collective behavior of brain tumor cells under the hypoxia
condition was studied in \cite{khan}.

A new therapeutic method, recently suggested
\cite{palti1,palti2,palti3} for non-invasive treatment of glioma -
brain cancer by a radio-frequency electric field, also opens new
directions of understanding of glioma development. A specific task
emerging here is whether this new medical technology is effective
against invasive cells with a high motility, when a switching
between migrating and proliferating phenotypes takes place. As
well known, one of the main features of malignant brain cancer is
the ability of tumor cells to invade the normal tissue away from
the multi-cell tumor core, causing treatment failure \cite{re}.
This problem relates to modelling of the dynamics of cancer glial
cells in heterogeneous media (as brain cancer is) in the presence
of a radio-frequency electric field, which acts as a tumor
treating field (TTF) \cite{palti1,palti2,palti3}. As reported,
this transcarnial treatment by the low-intensity (1-3 V/cm),
intermediate-frequency (100-200kHz) alternating electric field,
produced by electrode arrays applied to the scalp, destroys cancer
cells that undergoing to division, while normal tissue cells are
relatively not affected\footnote{An explanation of this phenomenon
in the electrostatic framework is vague, since for this weak RF
electric field of the order of 1V/cm, the inter-bridge voltage
between the daughter cells is of the order $10^{-3}$V that is less
than the voltage fluctuations related with the cell shape
fluctuations.}. An important result of this new technology
treatment is increasing the survival period in twice
\cite{palti3}\footnote{In the latest phase III trial study for TTF
treatment for glioblastoma \cite{stupp2}, the TTF treatment has
median survival of 6.6 months versus 6 months of the chemotherapy
treatment.} .

Therefore, an essential question is how the TTF affects aggressive
migrating cells in the outer-invasive region with a low-rate of
proliferation. To shed light on this situation, a simplified toy
model for glioma treatment by the TTF has been suggested in
\cite{iom2012,iom2013}, where a mathematical task of the
migration-proliferation dichotomy was formulated in the framework
CTRW \cite{shlesinger,klafter}. Note, that the simplest
mathematical realization of the CTRW mechanism of the
migration--proliferation dichotomy was introduced for a comb model
\cite{iom2006}. In the framework of this toy model, it was
possible to estimate the effectiveness of the TTF treatment in the
outer-invasive region of the tumor development \cite{iom2012}. It
has also been shown that while the TTF is highly effective in the
multi-cell tumor core, its action is ineffective in the presence
of the migration-proliferation dichotomy \cite{iom2012}. This
result is mainly based on the $1D$ consideration, where the
fractal cancer composite is embedded in the $1D$ space. In
reality, the situation is much more complicated, since the fractal
cancer composite in the outer-invasive region develops in the $3D$
space. As a result of this, the TTF efficiency depends on the
fractal dimension of the cancer composite in the outer-invasive
region. Therefore, a more realistic model to estimate a medical
effect of brain cancer (glioma) treatment by the RF electric field
is suggested \cite{iom2013}. This model is based on a construction
of a 3D comb model for the cancer cells, where the outer-invasive
region of glioma cancer is considered as a fractal composite
embedded in the 3D space. In the framework of this 3D model it was
shown that the efficiency of the medical treatment by the TTF
depends essentially on the mass fractal dimension $\frD$ of the
cancer in the outer-invasive region.

In this paper we follow a CTRW consideration, suggested in
\cite{iom2006}. Description of fractional kinetics of glioma
development under the TTF treatment in the framework of the one
dimensional (1D) and the three dimensional (3D) comb models show
that the efficiency of the medical treatment depends essentially
on the mass fractal dimension of the cancer in the outer invasive
zone \cite{iom2012,iom2013}. The aim of this research is
understanding both the role of the migration-proliferation
dichotomy in fractional cancer cell transport and its influence on
a therapeutic effect due to the TTF.

\section{Self-Entrapping by Fission as Fractional
Mechanism of Tumor Development} \label{sec:part2}

In this section we formulate the migration--proliferation
dichotomy in the framework of the CTRW. A simplified scheme of
cell dissemination through the vessel network was considered by
means of the following two steps \cite{iom2006,iom2005}. The first
step is a biological process of cell fission. The duration of this
stage is ${\cal T}_f$. The second process is cell transport itself
with duration ${\cal T}_t$. Therefore the cell dissemination is
approximately characterized by the fission time ${\cal T}_f $ and
the transport time ${\cal T}_t$. During the time scale ${\cal T}_f
$, the cells interact strongly with the environment and motility
of the cells is vanishingly small. The duration of ${\cal T}_f $
could be arbitrarily large. During the second time ${\cal T}_t$,
interaction between the cells is weak and motility of the cells
leads to cell invasion, which is a very complex process controlled
by matrix adhesion \cite{Giese1}. It involves several steps
including receptor-mediated adhesion of cells to extracellular
matrix (ECM), matrix degradation by tumor-secreted proteases
(proteolysis), detachment from ECM adhesion sites, and active
invasion into intercellular space created by protease degradation.
It is convenient to introduce a ``jump'' length $X_t$ of these
detachments as a distance which a cell travels during the time
${\cal T}_t$. Hence, the cells form an initial packet of free
spreading particles, and the contribution of cell dissemination to
the tumor development process consists of the following time
consequences:
\begin{equation}\label{tr2}
{\cal T}_f(1){\cal T}_t(2){\cal T}_f(3)\dots \, .
\end{equation}
There are different realizations of this chain of times, due to
different durations of ${\cal T}_f(i)$ and ${\cal T}_t(i)$, where
$i=1, 2, \dots$. Therefore, one concludes that transport is
characterized by random values ${\cal T}(i)$ which are waiting (or
self--entrapping) times between any two successive jumps of random
length $X(i)$. This phenomenon is known as a continuous time
random walk (CTRW) \cite{MW}. It arises as a result of a sequence
of independent identically distributed random waiting times ${\cal
T}(i)$, each having the same PDF $w(t), ~t>0$ with a mean
characteristic time $T$ and a sequence of independent identically
distributed random jumps, $x=X(i)$, each having the same PDF
$\lambda(x)$ with a jump length variance $\sigma^2$. It is worth
mentioning that a cell carries its own trap, by which it is set
apart from transport. This process of self--entrapping differs
from the standard CTRW, where traps are external with respect to
the transporting particles. The crucial point of the fractional
transport is the power law behavior of the waiting time PDF
\be\label{pdf_w} w(t)= \alpha\clT/(1+t/\clT)^{1+\alpha} \ee
where $0<\alpha<1$ and $\clT$ is a characteristic time. In this
case the averaged time is infinite. A proper explanation of eq.
(\ref{pdf_w}) can be the following quotation from Ref.
\cite{shlesinger}: ``A process with the long tailed pausing time
distribution would suffer a very sporadic behavior -- long
intermittencies may exist, followed by bursts of events. The more
probable pauses between events would be short but occasionally
very long pauses would exist. Given a long pause, there is still a
smaller but finite probability that an even longer one will occur.
It is on this basis that one would not be able to measure a mean
pausing time by examining data.'' Some justification of eq.
(\ref{pdf_w}) for the fission times can be presented by proposing
multi-time scales of self--entrapping. We can consider that
self-entrapping for different generations of cells has different
mean characteristic time scales, see Appendix A.  One obtains that
the PDF, which accounts for all exit events from proliferation
occurring on all time scales, has the power law asymptotic of eq.
(\ref{pdf_w}). Obtained distribution of eq. (\ref{pdf_w}) is
valid, when cell transport is considered on a fractional
subdiffusive structure such as a comb model.

\section{Comb-Like Model with Proliferation }
\label{sec:part3}

Fractional transport of cells, namely subdiffusion, can be
described in the framework of the comb model \cite{em1}. The comb
model is an example of subdiffusive 1d media where CTRW takes
place  along the $x$ structure axis. Diffusion in the $y$
direction plays the role of traps with the PDF of delay times of
the form $w(t)\sim1/(1+t/\clT)^{3/2}$. A special behavior of
diffusion on the comb structure is that the displacement in the
$x$--direction is possible only along the structure axis ($x$-axis
at $y=0$). Thus, the diffusion coefficient in the $x$--direction
is $D_{xx}=D\delta(y)$, while the diffusion coefficient in the
transversal $y$--direction is a constant $D_{yy}=D_0$. A random
walk on the comb structure is described by the distribution
function $P=P(x,y,t)$ and the current
\[{\bf j}=(-\delta(y)D\frac{\prt P}{\prt x},\, -D_0\frac{\partial
P}{\partial y})\, .\] %
The continuity equation with proliferation $C(P)$ yields the
following Fokker--Planck equation
\begin{equation}\label{comb1}
\frac{\partial P}{\partial
t}-\delta(y)D\frac{\partial^2P}{\partial x^2}
-D_0\frac{\partial^2P}{\partial y^2}= C(P) \, ,
\end{equation}
where the diffusion coefficients can be related to the CTRW
parameters $D=\sigma^2/\clT$. The initial condition $
P_(x,y,0)=P_0(x)\delta(y)$ is an initial distribution on the $x$
axis, and the boundary conditions are taken on infinities
$P(t)=P^{\prime}(t)=0 $ for both the $x$ and $y$ coordinates. The
primes denote the spatial derivatives.

It is convenient to work with dimensionless variables and
parameters. In the case of normal diffusion, when $D_x={\rm
const}$, the dimensionless time and coordinates are obtained by
re-scaling with relevant combinations of the comb parameters $D_x$
and $D_0$. One obtains the following dimension variables for time
$(D_0^3/D_x^2)t\rightarrow t$ and for the coordinates
$D_0x/D_x\rightarrow x,~~D_0y/D_x\rightarrow y$.

We consider a possible mechanism of tumor cell proliferation. The
term $C(P)$ in eq. (\ref{comb1}) determines the change in the
total number of transporting cells due to proliferation at rate
$\tilde{\cal C}$. This can be considered as a linear approximation
of a logistic population growth \cite{petrovskii}
\be\label{logistic} C(P)=\tilde{\clC} P(1-P/K)\, , \ee where $K$
is the carrying capacity of the environment (see {\em e.g.},
\cite{murray}). It is worth stressing that linearization is
important in the use of the powerful machinery of the Laplace
transform. When $P/K\rightarrow P<1/2$ and
$\clC=K\tilde{\clC}D_x^2/D_0^3$, then the linearization $C(P)=\clC
P$ is valid \cite{petrovskii}. In the opposite case, when
$P_1>1/2$ the growth is approximated by $C(P)=\clC\bar{P}$, where
$\bar{P}=1-P$. According to the migration--proliferation dichotomy
in the comb model, the transporting cells along the $x$ axis do
not proliferate. This means that cells proliferate only if they
have a non--zero $y$ coordinate. Therefore,
$C(P)=\clC(1-\delta(y))P$, and eq. (\ref{comb1}) reads in the
dimensionless form
\begin{equation}\label{comb1_prol1}
\frac{\partial P}{\partial t}-
\delta(y)\frac{\partial^2P}{\partial x^2}
-\frac{\partial^2P}{\partial y^2}= \clC\left(1-\delta(y)\right)P
\, .
\end{equation}
When $\clC>0$, eq. (\ref{comb1_prol1}) describes cell transport
with proliferation, and the PDF $P$ corresponds to a low
concentration of cells. In the opposite case, when $\clC<0$, eq.
(\ref{comb1_prol1}) describes fractional cell transport with
degradation that corresponds to a high cell concentration, and $P$
exchanges for $\bar{P}$.

The first term in the r.h.s. of eq. (\ref{comb1_prol1}) is
eliminated by substitution $P=e^{\clC t}F$. Carrying out the
Laplace transform $\tilde{F}(s,x,y)=\hat{\clL}[F(x,y,t)]$ and
looking for the solution in the form
$\tilde{F}=e^{-\sqrt{s}|y|}f(x,s)$, one obtains
\begin{equation}\label{pat4}
 F(x,y,t)=\hat{\clL}^{-1}\left[f(x,s)\exp(-\sqrt{s}|y|)\right] \, .
\end{equation}
As admitted, the true motion is in the $x$ axis, while the $y$
axis is an auxiliary, and integration over $y$ is performed.
Integrating eq. (\ref{comb1_prol1}) with respect to the variable
$y$ and introducing the PDF
\be\label{frac1} P_1(x,t)=\int_{-\infty}^{\infty}P(x,y,t)dy \, ,
\ee %
one obtains the following equation for $F_1=e^{-\clC t}P_1$ in the
Laplace space $\tilde{F}_1(s)=\hat{\clL}[F_1(t)]$:
\begin{equation} \label{pat5}
s\tilde{F}_1-\prt_x^2f=P_0(x)-\clC f \, .
\end{equation}
Integrating eq. (\ref{pat4}) over $y$, we obtain a relation
between the PDFs of the total number of cells $F_1$ and
transporting number of cells $f$ in the Laplace space
 \[f\equiv\tilde{F}(x,y=0,s)=(1/2)\sqrt{s}\tilde{F}_1(x,s)\, .\]
Substitution of this relation in eq. (\ref{pat5}) yields, after
the Laplace inversion, the Fokker--Planck equation for the
distribution $F$. To this end, eq. (\ref{pat5}) is multiplied by
$\sqrt{s}$ and then by virtue of eq. (\ref{B6}) the inverse
Laplace transform yields the following equation for $F_1$
\be\label{pat8} %
2D_C^{1/2}F_1-\prt_x^2F_1=-\clC F_1  \, , \ee where $D_C^{\alpha}$
is the fractional derivative in the Caputo form
\cite{podlubny,SKM} (see Appendix C). This equation describes
fractional transport of cells with fission when $\clC>0$ and
degradation when $\clC<0$, where the sign of $\clC$ depends on
either $P=e^{\clC t}F<1/2$, or $P>1/2$ \footnote{Since
$\partial\bar{P}=-\partial P$, eq. (\ref{comb1_prol1}) for $P$
(when $P<1/2$) just coincides with one for $\bar{P}=1-P$ (when
$P>1/2$). The only difference is when $P<1/2$, $\clC>0$, while for
$P>1/2$ one has $\clC<0$.}.

\section{Fractional Dynamics of Untreated Cancer}\label{sec:part4}

As  shown, the cell fission is a source of the fractional time
derivatives. This equation can be extended for an arbitrary
fractional exponent $0<\alpha<1$: $1/2\rightarrow \alpha$.
Therefore, this generalization of eq. (\ref{pat8}) yields
\be\label{pat8_a} %
D_C^{\alpha}F_1-\alpha\prt_x^2F_1=-\alpha\clC F_1
 \, .
\ee %
Taking into account that $D_C^{\alpha}$ can be expressed by the
Riemann--Liouville fractional derivatives $D_{RL}^{\alpha}$ (see
Appendix C) $D_C^{\alpha}=D_{RL}^{\alpha-1}D_{RL}^{1}$ and
$D_{RL}^{\alpha-}D_{RL}^{1-\alpha}=1$, we obtain another,
standard, form for the fractional Fokker--Planck equation (FFPE)
with proliferation, or degradation, %
\be\label{pat8_b} \frac{\prt F_1}{\prt t}- \alpha
D_{RL}^{1-\alpha}\frac{\prt^2F_1}{\prt x^2}= -\alpha\clC
D_{RL}^{1-\alpha}F_1\, . \ee %
To solve eq. (\ref{pat8_b}), we use the separation of variables
\cite{klafter}. We consider an analytical solution for the $P<1/2$
using the following substitution
\be\label{anzatz1}
F_1(x,t)=\sum_nT_n(t)\phi_n(x)\, . \ee %
Therefore, a solution which corresponds to the initial condition
$P_0(x)$, is determined by the Green function $G(x,t|x',0)$:
\be\label{PWP}  F_1(x,t) =
\int_{-\infty}^{\infty}dx'G(x,t|x',0)P_0(x')  =
\int_{-\infty}^{\infty}dx' \int dk T_k(t)\phi_k(x)
\phi_k^*(x')P_0(x')\, . \ee %
Here $ \phi_k(x)$ is a solution of the eigenvalue problem
$$-\frac{\prt^2\phi_k}{\prt x^2}=\lambda(k)\phi_k\, , $$
where $\lambda(k)=k^2$ is the continuous spectrum with
eigenfunctions
\be\label{eigfun} %
\phi_k(x)=\exp\Bigl[\pm kx\Bigr]\, . \ee %
The temporal eigenfunction $T_k(t)$ is governed by the fractional
equation
\be\label{fraceq}
\dot{T}_k(t)+\alpha\lambda_{\clC}(k)D_{RL}^{1-\alpha}T_k(t)=0\, ,
\ee %
where $\lambda_{\clC}(k)=(k^2+\clC)$. The solution is described by
the Mittag--Leffler function $E_{\alpha}(z)\equiv E_{\alpha,1}(z)$
\cite{batmen} (see Appendix C)
\be\label{solT}
T_k(t)=E_{\alpha}\left[\alpha\lambda_{\clC}(k)t^{\alpha}\right] \,
, \ee %
where $T_k(0)=1$, and $E_{\alpha}(z)$ has the initial
stretched exponent behavior \be\label{solT_a}
T_k(t)\sim\exp\left[-[\alpha\lambda_{\clC}(k)t^{\alpha}/\Gamma(1+\alpha)\right]
\ee which turns over to the power law long--time asymptotics
\be\label{solT_b} T_k(t)\sim
\left[\Gamma(1-\alpha)\alpha\lambda_{\clC}(k)t^{\alpha}
\right]^{-1} \, . \ee %
Using these properties of $E_{\alpha}(z)$, the fractional
spreading of cancer cells can be evaluated analytically for both
initial and long--time behaviors. Substitution of eqs.
(\ref{eigfun}) and (\ref{solT_a}) in eq. (\ref{PWP}) yields the
following initial time solution
\bea\label{solini} & P_1(x,t)\propto
\sqrt{\frac{\pi\Gamma(1+\alpha)}{\alpha t^{\alpha}}}
\exp\left[\clC t-\alpha\clC t^{\alpha}/\Gamma(1+\alpha)\right]
\nonumber \\
& \times \exp\left[-\Gamma(1+\alpha)x^2/4\alpha t^{\alpha}\right]
\, . \eea %
Analogously, the long--time solution is
\be\label{sollong} P_1(x,t)\propto\frac{1}{\alpha
t^{\alpha}\Gamma(1-\alpha)} \exp\left[\clC
t-\sqrt{\clC}|x|\right]\, , \ee %
where we take, for clarity, $P_0=\delta(x)$ for both the short and
long time solutions. These two solutions (\ref{solini}) and
(\ref{sollong}) corresponds to different scales. Solution of eq.
(\ref{sollong}) describes long-time/sort-scale dynamics. When the
argument in the exponential function is zero, it corresponds to
the front of cell invasion with equation $x\sim l_0=\sqrt{\clC}t$.
This is a so-called linear model which describes a solid tumor
growth. In this region with $x<l_0$ the exponential growth
$e^{\clC t}$ is dominant. Sundiffusion described by eq.
(\ref{solini}) corresponds to the cell transport in the
outer-invasive zone with $x>l_0$. When $\clC\rightarrow 0$ only
this solution takes place. Therefore, we have the the cell
spreading in the core region with $x<l_0$ is due to the cell
proliferation, while in the outer-invasive zone the cell motility
is the main engine of the cell spreading.

\section{Cell Kinetics in Presence of the TTF}\label{sec:part5}

Let us consider cell kinetics in the outer-invasive zone in more
detail. To this end we consider the fractal cancer development in
the presence of the TTF. This process can be described by
fractional kinetics in the framework of the comb model, as well,
where it is easier to draw an intelligible picture of interplay
between high-motility of aggressive cancer cells and the TTF in
the outer-invasive region. Contrary to the $1D$ comb model, in
this section we extend our consideration of the treated cancer to
the three dimensional cancer development, where proliferation
takes place inside a fractal composite, embedded in the $3D$ space
with the fractal dimension $\frD<3$.

In the $3D$ comb model, this anomalous diffusion can be described
by the $4D$ distribution function $P=P(\bfx,y,t)$, and by analogy
with the $1D$ comb model (\ref{comb1_prol1}), a special behavior
here is the displacement in the $3D$ $x$--space at $y=0$. The
Fokker-Planck equation in the same dimensionless variables reads
\be\label{fck_1}  %
\prt_tP= \delta(y) \Delta P +d\prt_y^2P\, , \ee %
where $d$ is an effective diffusion coefficient and
$\Delta=\sum_{j=1}^3\prt_{x_j}^2$.

\subsection{Comb Model with Proliferation and TTF}

Obviously, cell fission/division is random in the $x$--space and
discontinuous, contrary to that in the tumor core. Therefore, the
outer-invasive region of the cancer can be reasonably considered
as a random fractal set $F_{\frD}(\bfx)=F_{\alpha}(x_1)\times
F_{\beta}(x_2)\times F_{\gamma}(x_3)$ embedded in the $3D$ space,
as, for example, for low-grade astrocytomas \cite{Giese2}, with
the fractal dimension, $0<\frD<3$ and $\alpha+\beta+\gamma=\frD$.
For simplicity, we take $\alpha=\beta=\gamma=\frD/3=\nu$.

The effective diffusion coefficient in eq. (\ref{fck_1}) becomes
inhomogeneous $d\rightarrow d\chi(\bfx)$, where
$\chi(\bfx)=\chi(x_1)\chi(x_2)\chi(x_3)$ is a characteristic
function of the fractal, such that $\chi(x_j)=1$ for $x_j\in
F_{\frD}(\bfx)$ and $\chi(x_j)=0$ for $x_j\notin F_{\frD}(\bfx)$,
where $x_j$ are the Cartesian coordinates $j=1,2,3$. Now we take
into account the influence of the TTF that affects (destroys) only
quiescent cells, belonged to the proliferation phenotype,
according to Refs.~\cite{palti1,palti2,palti3}. Mathematically,
this process is expressed by diffusion in the $y$ direction with
decay:
\be\label{fck_2}  %
d\frac{\prt^2P(\bfx,y,t)}{\prt y^2}\Rightarrow
\chi(\bfx)[d\frac{\prt^2}{\prt y^2}-C]P(\bfx,y,t)\, , \ee   %
where coefficient $C$ defines a difference between the
proliferation and the degradation rate. In general case, $C$ is a
random function of time and space. For example, it was considered
as a random death rate for the random walk in the discrete
inhomogeneous media \cite{fiz2012}. Here we take it as a positive
averaged constant value.

Summarizing these arguments, mapping the glioma problem onto the
$4D$ comb model can be described by the following rules.
(\textit{i}) The dynamics of cancer cells takes place in the $3D$
space, which is described by three $x$ coordinates
$(x_1.x_2,x_3)$. (\textit{ii}) The $y$ axis corresponds to a
supplementary coordinate that introduces the
migration-proliferation dichotomy for the model. Therefore, at
$y=0$ the cells migrates and are not affected by the TTF.
Contrarily, the cells with $y\neq 0$ proliferate and are subjected
to the TTF.

Taking this into account, one arrives at the equation of  the
cancer development in the presence of the TTF
\be\label{fsk_3}  %
\frac{\prt P}{\prt t}=\delta(y)\Delta
P+\chi(\bfx)[d\frac{\prt^2}{\prt y^2}-C]P(\bfx,y,t)\, .
\ee %
First, we apply the Fourier transform to eq. (\ref{fsk_3}) with
respect to the $x_j$ coordinates. To this end, we rewrite eq.
(\ref{fsk_3}) in the form of convolution integrals.

Therefore, as shown in \cite{iom2013} and in Appendix B, fractal
cancer development in the presence of the TTF can be considered as
a random fractal composite of cancer cells embedded in the 3D.
Following coarse graining and averaging procedure, described in
Appendix B, we arrive at the $3D$ comb model that describes the
fractal cancer development in the outer-invasive region of glioma
in the presence of the TTF
\be\label{fsk_7}  %
\prt_tP(r,y,t)=\delta(y)\Delta
P(r,y,t)+[d\prt_y^2-C](-\Delta)^{\frac{3-\frD}{2}}P(r,y,t)\, , \ee
where $P(r,y,t)$ is the radial function in the 3D $x$ space
$r=|\bfx|$.

\subsection{ Dynamics in the Fourier-Laplace Space }

Equation (\ref{fsk_7}) can be considered as a starting point of
the analysis, and its solution will be obtained by means of the
Fourier and the Laplace transforms. Performing the Fourier
transform, constructed in the Appendix B, one obtains eq.
(\ref{fsk_7}) in the Fourier space
\be\label{fsk_8}  %
\prt_t\bar{P}=-k^2\delta(y)\hat{P}+
k^{3-\frD}[d\prt_y^2\hat{P}-C\bar{P}]\, . \ee   %
The last term in the r.h.s. of eq. (\ref{fsk_8}) is eliminated by
the substitution
\be\label{fsk_9}  %
\bar{P}(k,y,t)=\exp(-Ck^{3-\frD}t)F(k,y,t)=e^{-Ck^{-\alpha}t}F(k,y,t)\, , \ee   %
where $\alpha=\frD-3$.

The next step of the analysis is the Laplace transform  in the
time domain $$\hat{\clL}[F(k,y,t)]=\tilde{F}(k,y,s)\, .$$ Looking
for the solution of the Laplace image in the form
\be\label{fsk_10}  %
\tilde{F}(k,y,s)=\exp[-|y|\sqrt{k^{\alpha}s/d}]f(k,s)\, ,  \ee  %
one arrives at the intermediate expression in the form of the
Laplace and Fourier inversions
\be\label{fsk_11}  %
P(r,y,t)=\hat{\clF}_k^{-1}\left\{\exp(-Ck^{-\alpha}t)\hat{\clL}_t^{-1}
\left[ \frac{e^{-|y|\sqrt{sk^{\alpha}/d}}}{2\sqrt{sd
k^{-\alpha}}+ k^2}\right]\right\}\, . \ee %

As admitted above, the $y$ axis is the auxiliary, or supplementary
coordinate, which determines the cell proliferating process (cell
fission). Therefore to find the complete distribution of cancer
cells in the $x$ space, integration over $y$ is performed (see
Sec. \ref{sec:part4}):
\be\label{fsk_12} %
\overline{P}(r,t)=\int_{-\infty}^{\infty}P(r,y,t)dy\, . \ee   %
Both the integration over $y$ and the inverse Laplace transform
are carried out exactly. This, eventually, yields a solution in
the form the $3D$ Fourier inversion
\be\label{fsk_13} %
\overline{P}(r,t)=\frac{1}{(2\pi)^3}\int_{-\infty}^{\infty}e^{-i\bfk\cdot\bfx}
\exp(-Ck^{3-\frD}t)\clE_{\frac{1}{2}}\Big(-\frac{1}{2}\sqrt{k^{1+\frD}t/d}\Big)d^3k
\, . \ee %
Here $$\clE_{\alpha}(-z)=\frac{1}{2\pi
i}\int_{\gamma}\frac{u^{\alpha-1}e^udu}{u^{\alpha}+z}$$ is the
Mittag-Leffler function defined by the inverse Laplace transform
with a corresponding deformation of the contour of the integration
\cite{batmen}.

\subsection{True Distributions}

Solution (\ref{fsk_13}) is a convolution of the kernel of the TTF
treatment $\clR(z)$ and the untreated cancer distribution
$\clP(z)$
\be\label{fsk_14}  %
\overline{P}(r,t)=\clR\star\clP\, .  \ee  %
When $C=0$, which means that the TTF compensates proliferation,
the solution is described by the Mittag-Leffler function with the
scaling variable $z=r/t^{\frac{1}{1+\frD}}$. This scaling
determines the cancer cell expansion
\be\label{fsk_15} %
r\sim t^{\frac{1}{1+\frD}} \ee %
that depends essentially  on the fractal dimension of the
proliferation volume of the fractal cancer composite and reflects
the migration-proliferation dichotomy in the outer-invasive
region. Indeed, for the fractal cancer volume (or mass)
$\mu(r)\sim r^{\frD}$, the cancer development is superdiffusive
when $\frD<1$, while for $\frD>1$ the latter spreads
subdiffusively. This property is pure kinetic and, apparently, is
universal for the cancer development and related to the fractal
dimension of the cancer \cite{breasrcancer}.

Now let us return to the convolution integral (\ref{fsk_14}).
To avoid awkward expressions of integrations with the
hypergeometric functions, we consider particular cases of the
fractal dimension $\frD=2$ and $\frD=1$. For $\frD=2$, due to the
scaling argument, one obtains that the untreated cancer spreads
subdiffusively $\lgl r^2\rgl\sim t^{\frac{2}{3}}$, while for the
TTF kernel we have
\be\label{fsk_16}  %
\clR(r,t)=\frac{1}{\sqrt{(2\pi)^3r}}\int_0^{\infty}
e^{-Ctk}k^{\frac{3}{2}}J_{\frac{1}{2}}(kr)dk=
\frac{1}{3\pi^2(Ct)^3}{}_2F_1(\frac{3}{2},2;\frac{3}{2};-\frac{r^2}{(Ct)^2})\,
.  \ee  %
Here ${}_2F_1(a,b;c;z)$ is the hypergeometric function. This
yields the power law decay of the distribution function
\be\label{fsk_17}  %
\clR(r,t)=\frac{(Ct)^{-3}\pi^{-2}}{3[1+r^2/(Ct)^2]^2}\,
.\ee  %
This power law kernel shows that the TTF is inefficient for
$\frD>1$ in the presence of the migration proliferation dichotomy.
It is tempting to calculate the second moment $\lgl r^2(t)\rgl$
with the distribution $\overline{P}(r,t)$. In this case, one
should recognize that a cutoff $r=t$ of the L\'{e}vy flights for
$r,t\gg 1$ should be performed. This is a well known procedure
\cite{zumofen}, used for the L\'{e}vy walks.

\subsubsection{$\frD=1$}

The situation changes when $\frD\leq 1$. In this case the TTF
leads to the Brown exponential cutoff of the cancer spread in eq.
(\ref{fsk_14}). For $\frD=1$ the problem is analytically
treatable. For the small argument, which corresponds (for a short
time) to a long-scale tail of the distribution, the Mittag-Leffler
function decays exponentially
$\exp\Big(-K_{\frac{1}{2}}\sqrt{|k|^{1+\frD}t}\Big)$
\cite{klafter,batmen} with the generalized transport coefficient
$K_{\frac{1}{2}}=[2\Gamma(3/2)\sqrt{d}]^{-1}$. This yields the
solution for the compensated cancer with $C=0$ in the form of the
hypergeometric functions like in eqs. (\ref{fsk_16}) and
(\ref{fsk_17}). Following \cite{prudnikov}, one obtains
\be\label{fsk_18}  %
\clP(r,t)=\frac{K_{\frac{1}{2}}\sqrt{t}}
{(2\pi)^3(1+r^2/K_{\frac{1}{2}}^2t)^2}
\left[K_{\frac{1}{2}}\sqrt{t}+
\sqrt{K_{\frac{1}{2}}^2t+r^2}\right]^{-\frac{1}{2}}\, .
\ee  %
This metastatic power law behavior is restricted by the Brown
distribution due the TTF kernel
\be\label{fsk_19}  %
\clR(r,t)=\hat{\clF}^{-1}[e^{-Ck^2t}]=\frac{1}{(4\pi
Ct)^{3/2}}\exp\left(-\frac{r^2}{4Ct}\right)\, .  \ee %
The second moment is a good characteristic to show the TTF
influence. One obtains from eq. (\ref{fsk_18}) for the compensated
cancer $\lgl r^2\rgl\sim t^{\frac{3}{2}}$ for $r\gg 1$ that
corresponds to superdiffusion at the large scale asymptotics, and
the cutoff at $r=t$ is taken into account. The same calculation
with the TTF kernel yields an effective treatment with $\lgl
r^2\rgl\sim t^{\frac{3}{4}}$ that corresponds to the
superdiffusion--subdiffion transition due to the TTF. Obviously,
that untreated cancer with $C<0$ leads to the exponential
spreading of cancer cells due to the exponential proliferation.

\subsection{Numerical Estimations of Eq. (\ref{fsk_13})}

As shown in eqs. (\ref{fsk_17}) and (\ref{fsk_19}) analytical form
of the TTF operator depends on fractal dimension $\frD$. Since
analytical estimation of eq. (\ref{fsk_13}) leads to awkward
expressions of integrations with the hypergeometric functions,
numerical procedure is performed. The results are depicted in
Fig.~1 for different values of the fractal dimension $\frD$.

\begin{figure}[ph]
\centerline{\epsfig{file=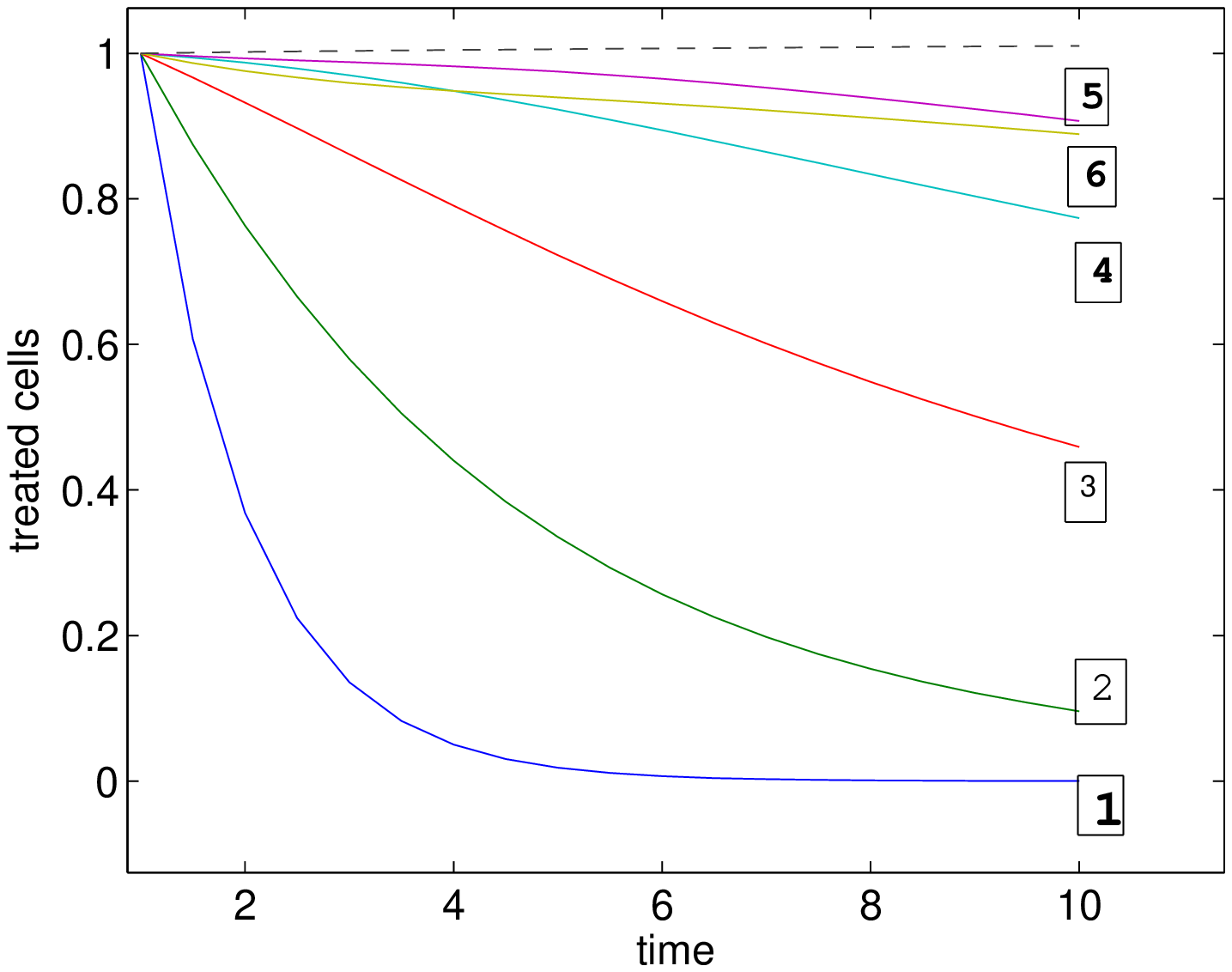,width=8cm}} \vspace*{8pt}
\caption{Dynamics of the treated cancer distribution for different
values of the fractal dimension $\frD$, where from plots from 1 to
6 correspond to $\frD=(3\, ,2.5\, ,2\, ,1.5\, ,1\, ,0.5)$ }
\end{figure}
As obtained, the maximal therapeutic effect takes place at
$\frD=3$ that immediately follows from eq. (\ref{fsk_13}). One
should recognize that the solution for the compensated cancer
$\clP(r,t)$ is exactly the form of the interplay between the TTF
and subdiffusion, which is the result of the
migration-proliferation dichotomy. Another manifestation of this
interplay is the fractal dimension $\frD<3$ that leads to
metastatic behavior of either the TTF kernel or compensated cancer
solution.

\section{Conclusion}

The present study focuses on the influence of cell proliferation
on transport properties. The mathematical formulation of this
proliferation-migration dichotomy is based on the two main stages:
cell fission with the self-entrapping time ${\cal T}_f$ and cell
transport with durations ${\cal T}_t$. By virtue of these two time
scales a description of tumor development is reduced to a CTRW
process. A toy model of cancer development is suggested by using
heuristic arguments on the relation between tumor development and
the CTRW. In this case a fractional tumor development becomes a
well defined problem since a mathematical apparatus of CTRW is
well established (see  {\it e.g.}
\cite{klafter,shlesinger,podlubny,SokKlafBlum}). The constructed
model is a modification of a so--called comb structure \cite{em1}.
An important feature of this consideration of cell transport in
the framework of the comb model is an essential enhancement of
anomalous transport due to proliferation. Moreover, we obtained
that the distribution function of the fractional transport depends
on only two parameters, namely, scaled proliferation rate $\clC$
and the fractional exponent $\alpha$, where $\alpha=1/2$ for the
comb model.

The next step is studying glioma cancer development in the
outer-invasive region in the presence of a tumor treating field.
The model is based on a construction of a 3D comb model for the
cancer cell transport, where the outer-invasive region of glioma
cancer is considered as a fractal composite, embedded in the 3D
host of the normal cell tissues. The description is performed in
the four-dimensional $(\bfx,y)$ space, where the real
three-dimensional $\bfx$ space stands for the description of real
cancer development, while the supplementary $y$-coordinate is
introduced to described a non-Markovian process in the framework
of the Markovian description. From the biological point of view,
this corresponds to the migration-proliferation dichotomy of the
cancer cells, where the influence of the TTF is considered, as
well. Therefore, the kinetic equation (\ref{fsk_7}) in the
$(\bfx,y)$ space is constructed by means of the coarse-graining,
or embedding procedure inside the fractal space. This corresponds
to the averaging in the 3D Fourier space in eq. (\ref{fsk_6}), and
can be (roughly) considered as a generalization of the $1D$
procedure, based on averaging extensive physical values and
expressed by means of a smooth function over a Cantor set that,
eventually, leads to fractional integration \cite{NIG92,NIG98}.

The efficiency of the TTF is estimated in the form of the
convolution in eqs. (\ref{fsk_13}) and (\ref{fsk_14}). This
expressions describe the influence of the TTF on the cancer
development. The efficiency of the medical treatment by the TTF
depends essentially on the fractal dimension $\frD$, and the TTF
is the most efficient for $\frD=3$. But, in reality, the
outer-invasive zone is a fractal composite with the fractal
dimension $\frD<3$.

Another result relates to the spread of the compensated cancer,
which is determined by eq. (\ref{fsk_15}). This result reflects
the migration-proliferation dichotomy, namely the dependence of
the cancer cells spread on the fractal dimension of the
proliferation volume. For example, the cancer development is
superdiffusive when $\frD<1$, while for $\frD>1$ it spreads
subdiffusively. This property is pure kinetic and, apparently, is
universal for the cancer development with a variety of
bio-chemical processes. Recently an experimental validation of
this kind phenomenon has been obtained for the metastatic
detaching under \textit{in vitro} studying the breast cancer
\cite{breastcancer}. Important question here also is aging the
treatment. Initial times of the cancer development and treatment
are different, and the time difference is unknown. Moreover, since
the TTF acts on the proliferating cells only, the migration
proliferation dichotomy leads to a particular case of a more
general problem of aging population splitting \cite{schulz}. For
the present analysis this general approach \cite{schulz,barkai}
can be important for understanding the efficiency of the TTF. This
can be an interesting issue for future studies.

It is also worth noting that a general solution in the form of the
convolution $\overline{P}(r,t)=\clR\star\clP$ makes it possible to
consider the compensated cancer in a more general framework of the
fractional Fokker-Planck equation, namely \cite{iom2011}
\[\prt_t^{p}\clP=(-\Delta)^{\frac{q}{2}}\clP\, , \]
where Caputo fractional derivative $\prt_t^{p}$ is responsible for
the migration-proliferation dichotomy, while $q=q(\frD)$ reflects
the fractal dimension of the tumor development in the
outer-invasive region. In our case, $p=\frac{1}{2}$, according the
comb model construction, while $q=(1+\frD)/4$ is an universal
parameter, which determines the fractional space derivative due to
the fractal dimension of the quiescent/proliferating cancer cells.
In general case, $p\in (0,1)$, and it is determined by another way
of the introduction of the supplementary variable $y$
\cite{fiz2012,Cox}.

In conclusion, we discuss briefly a possible direct experiment,
confirming the existence of a fractal cancer composite in the
outer-invasive region. Cancer was considered as a fractal
composite where a random fractal inclusion of the cancer cells
$F_{\frD}(\bfx)=F_{\alpha}(x_1)\times F_{\beta}(x_2)\times
F_{\gamma}(x_3)$ is embedded in the $3D$ space of normal tissue
cells. Therefore, in the presence of the TTF, one can consider the
frequency-dependent permittivities of migrating cancer cells
$\vep_m$ and the normal tissue cells $\vep_n$ \cite{iom2012}.
Under certain frequency of the TTF, the condition $\vep_m<\vep_n$
can be fulfilled. These permittivities were observed in time
domain dielectric spectroscopy in experimental studies of the
static and dynamic dielectric properties of normal, transformed,
and malignant B- and T-lymphocytes \cite{feldman}. The solution of
the Maxwell equation for the electrostatic field in the frequency
domain yields an essential enhancement of the respond field inside
the random fractal dielectric composite of cancer cells.
Therefore, the respond electric field can be large enough to break
the cell membrane. For example, as shown in \cite{iom2012} the
electric field response can be of the order of  $10^4\div 10^5$
V/cm, which exerts the irreversible electroporation
\cite{Rubinsky} due to the external TTF with amplitude $\sim
1$V/cm. This can be a mechanism for ablation of cancer cells,
which effectively acts on migratory cancer cells.

A key quantity of this cancer treatment is localization of the
electroporation field inside the cancer. There is a
straightforward analogy with nanoplasmonics (see \textit{e.g.},
\cite{Shalaev,Stockman}), where the electric field enhancement is
due to a so-called surface-plasmon resonance for a
metal-dielectric composite, and localized surface plasmon
oscillations are charge density oscillations confined to the
conducting fractal nanostructure. The essential difference  is
that this biological cell enhancement of the electric field is not
resonant, but geometrical due to the fractal cancer structure
\cite{bi2011b}. In this connection, \textit{in vitro} experiments
can be important for further understanding the interplay between
the TTF and the migration-proliferation dichotomy. Theoretical
description of this phenomenon needs more realistic assumptions
than those suggested here in the framework of the comb model. Such
studies should be performed in the framework of more sophisticated
models of the switching between the migration and proliferation
phenotypes
\cite{Khain,Deutsch,Chauviere,kolobov,fir2011,fi07,fi08}, and the
next step is understanding how dielectric properties of cells
correlate with cell motility and cell fission. Such experimental
studies of the glioma cells can not be overestimated.

\section*{Acknowledgments}

This work was supported  by the Israel Science Foundation (ISF).

\appendix

\section{Power Law PDF}

\def\theequation{A. \arabic{equation}}
\setcounter{equation}{0}

As an example, we consider that  $j$-th generation of
self-entrapping is the Poisson process
$$w_j(t)=\tau_j^{-1}\exp\left(-t/\tau_j\right)$$
with the characteristic time scale $\tau_j=\tau^j$, where
$\tau=\tau_1=\clT$ is now an average time of cell divisions for
the first generation. Therefore, following
\cite{shlezinger1,blumen} and repeating exactly the analysis of
Ref. \cite{blumen}, we obtain, by taking into account events
occurring on all time scales, the following distribution:
$$w(t)=\frac{1-b}{b}\sum_{j=1}^{\infty}b^{j}\tau^{-j}
\exp\left(-t/\tau^l\right)\, ,$$ where $b<1$ is a normalization
constant. Therefore, the last expression is a normalized sum and
$$w(t/\tau)=\tau w(t)/b-(1-b) \exp\left(-t/\tau\right)/b\, .$$ Using
conditions $t\gg\tau>1/b$, one obtains that at longer times
$w(t/\tau)=\tau w(t)/b$. The last expression is equivalent to
\be\label{A1} w(t)\sim 1/t^{1+\alpha}\, , \ee  %
where $\alpha=\ln(b)/\ln(1/\tau)$.

\section{Coarse-Graining Procedure of Fractal Cancer Composite}

\def\theequation{B. \arabic{equation}}
\setcounter{equation}{0}

Using the auxiliary identity
$$\chi(x_j)f(x_j)\equiv\prt_{x_j}\int_{-\infty}^{x_j}\chi(y)f(y)dy\equiv
-\prt_{x_j}\int_{x_j}^{\infty}\chi(y)f(y)dy $$
with the boundary conditions $P(x_i=\pm\infty)=0$, this
integration with the characteristic function can be carried out by
means of a convolution \cite{iom2011,bi2011a,Ren}. Note, that
$$ \int_{-\infty}^{\infty}\chi(y)f(y)dy=\sum_{x_j\in
F_{\nu}}\int_{-\infty}^{\infty}f(y)\delta(y-x_j)dy\, , $$
where $\sum_{x_j\in F_{\nu}}\delta(y-x_j)=\mu^{\prime}(x)\sim
|x|^{\nu-1}$ is a fractal density, such that on the finite
interval $(-x,x)$, the integral
$$ \int_{-x}^xd\mu(y)\sim |x|^{\nu} $$
corresponds to the fractal volume. Therefore, due to Theorem $3.1$
in Ref.~\cite{Ren} we have
$$\int_0^xf(y)d\mu(y)\simeq
\frac{1}{\Gamma(\nu)}\int_0^x(x-y)^{\nu-1}f(y)dy\, , $$
which is defined for the finite fractal volume
$\mu(x)\equiv\mu(x_j)$.

In what follows we will use the terminology and useful notations
of fractional integration and differentiation
\cite{klafter,podlubny,SKM,SokKlafBlum}. Fractional integration of
the order of $\nu$ is defined by the operator (see Appendix C)
\be\label{fsk_4b1}   %
{}_{-\infty}I_x^{\nu}f(x)=
\frac{1}{\Gamma(\nu)}\int_{\infty}^xf(y)(x-y)^{\nu-1}dy\, , \ee %
\be\label{fsk_4b2} %
{}_xI_{\infty}^{\nu}f(x)=
\frac{1}{\Gamma(\nu)}\int_x^{\infty}f(y)(y-x)^{\nu-1}dy\, ,
\ee %
where $0<\nu<1$ and  $\Gamma(\nu)$ is the Gamma function. By means
of these fractional integration and differentiation (see Appendix
C)
\be\label{fsk_4a1} %
\clW_-^{1-\nu}f(x)= \prt_x[{}_{-\infty}I_x^{\nu}f(x)]\, , \ee%
\be\label{fsk_4a2} %
\clW_+^{1-\nu}f(x) =\prt_x[{}_xI_{\infty}^{\nu}f(x)]
\, , \ee %
one introduces the coarse-graining integration with the
characteristic function in the form of the Riesz fractional
derivative \cite{bi2011a}
\be\label{fsk_5}  %
\chi(x_j)P(x,y,t)\Rightarrow[\clW_-^{1-\nu}+\clW_+^{1-\nu}]P(x_j,y,t)
=\clW^{1-\nu}P(x_j,y,t)\,  .  \ee   %

\subsubsection{Random fractal composite}

We consider a random fractal with an averaged volume, embedded in
the $3D$, which is a function of a radius only $\mu(r)\sim
r^{\frD}$ \cite{berrypercival,benavraam}, where
$r=\sqrt{\sum_jx_j^2}$. Therefore, the distribution function and
the kernel of the fractional integration are the radial functions,
and therefore, fractional integrations over the Cartesian
coordinates $x_j$ are substituted by integrations over the radial
functions. This averaging procedure can be performed in the
Fourier space as follows
\be\label{fsk_6}  %
\hat{\clF}\left[\prod_jW^{1-\nu}P(r,y,t)\right]=
\prod_j|k_j|^{1-\nu}\hat{P}(\{k_j\},y,t) \Rightarrow
k^{3-\frD}\bar{P}(k,y,t)\, , \ee  %
where $k=\sqrt{\sum_j k_j^2}$ is the radius in the Fourier space
and $\bar{P}(k,y,t)=\hat{\clF}[P(r,y,t)]$. This averaging
substitute in the $3D$ Fourier space is an extension of $1D$
embedding in the fractal, obtained in \cite{Ren,NIG92}, in
agreement with Nigmatulin's arguments on a link between fractal
geometry and fractional integro-differentiation
\cite{NIG92,NIG98}.  This is constituted in the procedure of
averaging extensive physical values and expressed by means of a
smooth function over a Cantor set that, eventually, leads to
fractional integration \cite{NIG92,NIG98}.

Note, that we did not use here any property of the kernel as a
radial function that can be considered as the Riesz potential
\cite{SKM}, as well
\be\label{fsk_6a} %
\prod_j\prt_{|x_j|}\prod_j\frac{1}{|x_j-x_j^{\prime}|^{1-\nu}}\Rightarrow
\frac{1}{|x-x'|^{6-\frD}} =
\frac{\gamma(\alpha)}{\left(\sqrt{\sum_j(x_j-x_j^{\prime})^2}\right)^{3-\alpha}}\,
,\ee   %
where $\alpha=\frD-3$ and $\gamma(\alpha)\equiv\gamma(\frD)$ is
defined by Weber's integral
\be\label{fsk_6b} %
\int_0^{\infty}z^{\beta}J_{\nu}(z)dz=2^{\beta}
\Gamma\left(\frac{\nu+\beta+1}{2}\right)/\Gamma\left(\frac{\nu-\beta+1}{2}\right)
\ee  %
at the Fourier transform of the Riesz kernel
\be\label{fsk_6c}  %
\hat{\clF}[r^{\alpha-3}]=\frac{(2\pi)^{\frac{3}{2}}}{\sqrt{k}}\int_0^{\infty}
r^{\alpha-\frac{3}{2}}J_{\frac{1}{2}}(rk)dr=\gamma(\frD)k^{3-\frD}\,
.\ee  %
Following \cite{SKM} (ch. 25), we redefine $\prod_jW^{1-\nu}$ as
the fractional degree of the Laplace operator
$(-\Delta)^{-\alpha/2}$, namely
\be\label{fsk_6d}  %
\prod_jW^{1-\nu}P(r,y,t)\Rightarrow\frac{1}{\gamma(\alpha)}\int\frac{P(r')\prod_jdx_j'}
{|x-x'|^{3-\alpha}}\equiv
\left(-\Delta\right)^{-\frac{\alpha}{2}}P(r,y,t)\, .\ee  %
This yields (see also \cite{SKM})
\be\label{fsk_6e}  %
\hat{\clF}\left(-\Delta\right)^{-\frac{\alpha}{2}}P(r,y,t)=k^{-\alpha}\bar{P}(k,y,t)\,
.\ee  %
Eventually, we arrive at the $3D$ comb model (\ref{fsk_7})
\be\label{fsk_7a}  %
\prt_tP(r,y,t)=\delta(y)\Delta
P(r,y,t)+[d\prt_y^2-C](-\Delta)^{\frac{3-\frD}{2}}P(r,y,t)\, . \ee

\section{Fractional Integro--Differentiation}

\def\theequation{C. \arabic{equation}}
\setcounter{equation}{0}

Fractional integration of the order of $\alpha$ is defined by the
operator
\be\label{B1} %
{}_aI_x^{\alpha}f(x)=
\frac{1}{\Gamma(\alpha)}\int_a^xf(y)(x-y)^{\alpha-1}dy\, , \ee  %
where $\alpha>0,~x>a$ and  $\Gamma(z)$ is the Gamma function. The
fractional derivative is the inverse operator to
${}_aI_x^{\alpha}$ as $ {}_aD_x^{\alpha}f(x)={}_aI_x^{-\alpha}$
and ${}_aI_x^{\alpha}={}_aD_x^{-\alpha}$. Its explicit form is
\be\label{B2}  %
{}_aD_x^{-\alpha}=
\frac{1}{\Gamma(-\alpha)}\int_a^xf(y)(x-y)^{-1-\alpha}dy\, .\ee %
For arbitrary $\alpha>0$ this integral diverges, and as a result
of a regularization procedure, there are two alternative
definitions of ${}_aD_x^{-\alpha}$. For an integer $n$ defined as
$n-1<\alpha<n$, one obtains the Riemann-Liouville fractional
derivative of the form
\be\label{B3}  %
{}_aD_{RL}^{\alpha}f(x)=(d^n/x^n){}_aI_x^{n-\alpha}f(x)\, ,\ee  %
and fractional derivative in the Caputo form
\be\label{B4} %
{}_aD_{C}^{\alpha}f(x)= {}_aI_x^{n-\alpha}f^{(n)}(x)\, . \ee   %
When $a=-\infty$, the resulting Weyl derivative is
\be\label{B5}  %
\clW^{\alpha}\equiv{}_{-\infty}D_{W}^{\alpha}=
{}_{-\infty}D_{RL}^{\alpha}= {}_{-\infty}D_{C}^{\alpha}\, .\ee  %
One also has ${}_{-\infty}D_{W}^{\alpha}e^x=e^x$ This property is
convenient for the Fourier transform
$$\hat{\clF}\left[\clW^{\alpha}f(x)\right]=(ik)^{\alpha}\hat{f}(k)\,
, $$ where $\hat{\clF}[f(x)]=\hat{f}(k)$.

The Laplace transform can be obtained for Eq. (\ref{B4}). If
$\hat{L}f(t)=\tilde{f}(s)$ is the Laplace transform of $f(t)$,
then
\be\label{B6} \hat{L}[D_C^{\alpha}f(t)]=s^{\alpha}\tilde{f}(s)-
\sum_{k=0}^{n-1}f^{(k)}(0^+)s^{\alpha-1-k}\, . \ee %
We also note that
\be\label{B8}
D_{RL}^{\alpha}[1]=\frac{t^{-\alpha}}{\Gamma(1-\alpha)}\, , ~~
D_C^{\alpha}[1]=0\, . \ee %
The fractional derivative of a power function is
\be\label{B9}
D_{RL}^{\alpha}t^{\beta}=\frac{t^{\beta-\alpha}\Gamma(\beta+1)}
{\Gamma(\beta+1-\alpha)}\, , \ee %
where $\beta>-1$ and $\alpha>0$.
The fractional derivative from an exponential function can be
simply calculated as well by virtue of the Mittag--Leffler
function (see {\em e.g.}, \cite{podlubny}):
\be\label{B10} %
E_{\gamma,\delta}(z)=\sum_{k=0}^{\infty}
\frac{z^k}{\Gamma(\gamma k+\delta)} \, . \ee
Therefore, from Eqs.
(ref{B9}) and (\ref{B10}) we have the following expression
\be\label{B11} D_{RL}^{\alpha}e^{\lambda
t}=t^{\alpha}E_{1,1-\alpha}(\lambda t)\, . \ee




\begin{thebibliography}{00}  

\bibitem{MCB} H. Lodish, A. Berk, S.L. Zipursky, P. Matsudaira.
D. Baltimore, and J. Darnell, \textit{Molecular Cell Biology},
(W.H. Freeman and Company, New York, 2000).

\bibitem{AANS} AANS Classification of Brain Tumors,
http://www.aans.org/

\bibitem{hanahan} D. Hanahan and R. A. Weinberg, Cell \textbf{100}, 57
(2000).

\bibitem{stupp1} R. Stupp, W.P. Mason, and M.J, van den Bent, \textit{et al.},
N. Engl. J. Med. \textbf{352}, 987 (2005).

\bibitem{stupp2} R. Stupp, E.T. Wong, A.A. Kanner, textit{et al.},
Eur. J. Cancer  \textbf{48}, 2192 (2012).


\bibitem{Giese1} A. Giese \textit{et al.},
Int. J. Cancer \textbf{67}, 275 (1996).

\bibitem{Giese2} A. Giese \textit{et al.},
J. Clin. Oncology \textbf{21}, 1624 (2003).

\bibitem{Corcoran} A. Corcoran and R. F. Del Maestro,
Neurosurgery \textbf{53}, 174 (2003).

\bibitem{garay} T. Garay \textit{et al.},
Exper. Cell Res. In press (2013).

\bibitem{garay18} S. Khoshyomn, S. Lew, J. DeMattia, E.B. Singer,
and P.L. Penar,
J.  Neuro-Oncology \textbf{4}, 111 (1999).

\bibitem{garay26} A. Merzak, S. McCrea, S. Koocheckpour, and G.J.
Pilkington,
Br. J. Cancer \textbf{70}, 199 (1994).


\bibitem{garay37} M. Tamaki, \textit{et al.},
J. Neurosurg \textbf{87}, 602 (1997).

\bibitem{breastcancer} L. Jerby, L. Wolf, C. Denkert, G.Y. Stein,
\textit{et al.}, Cancer Res. doi:10.1158/0008-5472. CAN-12-2215.

\bibitem{Khain}
E. Khain E. and L.M. Sander, Phys. Rev. Lett. \textbf{96}, 188103
(2006).

\bibitem{Deutsch}
H. Hatzikirou, D. Basanta, M. Simon, K. Schaller, and A. Deutsch
Math. Med. Biol. \textbf{29}, 49 (2010).

\bibitem{Chauviere}
A. Chauviere, L. Prziosi, and H. Byrne,
Math. Med. Biol. \textbf{27,} 255 (2010).

\bibitem{kolobov}
A.V. Kolobov, V.V. Gubernov, and A.A. Polezhaev, Math. Model. Nat.
Phenom. \textbf{6}, 27 (2011).

\bibitem{fir2011}
S. Fedotov, A. Iomin, and L. Ryashko,
Phys. Rev. E \textbf{84}, 061131 (2011).
\bibitem{khain1} \newblock E. Khain, L. D. Sander, A. M.
Stein, 
\newblock Complexity \textbf{11}, 53 (2005).

\bibitem{athale} \newblock C. A. Athale, Y. Mansury, T.
S. Deisboeck,
\newblock J. Theor. Biol. \textbf{233}, 469 (2005).

\bibitem{Zhang1} L. Zhang, Z. Wang, J. Sagotsky, T.S. Deisboeck, J. Math.
Biol. \textbf{58,} 545 (2008).

\bibitem{Zhang2} L. Zhang, L. L. Chen, T. S. Deisboeck, Math. Comp. in
Simulation \textbf{79,} 2021 (2009).

\bibitem{iom2006} A. Iomin,
Phys. Rev. {\bf E 73}, 061918 (2006).

\bibitem{fi07} \newblock S. Fedotov and A. Iomin,
\newblock Phys. Rev. Lett. \textbf{98}, 118101 (2007).

\bibitem{fi08} S. Fedotov and A. Iomin,
\newblock Phys. Rev. E \textbf{77}, 031911 (2008).

\bibitem{Ch2} M. Tektonidis \textit{et al.}, J. Theor. Biology \textbf{287, }%
131 (2011).

\bibitem{khan} E. Khain \textit{et al., }Phys. Rev. E \textbf{83}, 031920
(2011).


\bibitem{palti1}
E.D. Kirson, \textit{et al}.,
Cancer Res. \textbf{64}, 3288 (2004).

\bibitem{palti2}
E.D. Kirson, \textit{et al}.,
Proc.Nat.Acad.Sci. USA \textbf{104}, 10152 (2007).

\bibitem{palti3} Y. Palti,
Europ. Oncological Disease \textbf{1}(1), 89 (2007).

\bibitem{re}
D. H. Geho \textit{et al.}, Physiology, \textbf{20}, 194 (2005).

\bibitem{iom2012} A. Iomin, Eur. Phys. J. E \textbf{35}, 42
(2012).

\bibitem{iom2013} A. Iomin,
Eur. Phys. J. Special Topics \textbf{222}, 1873 (2013).

\bibitem{shlesinger}  E.W. Montroll and M.F. Shlesinger,
in J. Lebowitz and E.W. Montroll  (eds) \textit{Studies in
Statistical Mechanics}, v. 11 (Noth--Holland, Amsterdam, 1984).

\bibitem{klafter} R. Metzler and J. Klafter,  Phys. Rep. \textbf{339}, 1 (2000).

\bibitem{iom2005}  A. Iomin, J. Phys.: Conference Series {\bf 7},
57 (2005); WSEAS Trans. Biol. Biomed. {\bf 2}, 82 (2005).

\bibitem{MW} E.W. Montroll and G.H. Weiss, J. Math. Phys. {\bf 6}, 167
(1965).


\bibitem{em1} G.H. Weiss and S. Havlin, Physica A {\bf 134}, 474 (1986).




\bibitem{petrovskii} S.V. Petrovskii and B.-L. Li, {\em Exactly Solvable
Models of Biological Invasion}, (Chapman \& Hall, Boca Raton,
2005).

\bibitem{murray} J.D. Murray, {\em Mathematical Biology},
(Springer, Heidelberg, 1993).


\bibitem{podlubny} I. Podlubny, \textit{Fractional Differential
Equations} (Academic Press, San Diego, 1999).


\bibitem{SKM} S.G. Samko, A.A. Kilbas, and O.I. Marichev,
\textit{Fractional Integrals and Derivatives} (Gordon and Breach,
New York, 1993).

\bibitem{batmen} H. Bateman and A. Erd\'elyi,
\textit{Higher Transcendental functions} (Mc Graw-Hill, New York,
1955), V. 3.

\bibitem{fiz2012} S. Fedotov, A. O. Ivanov, and A. Y. Zubarev,
Non-homogeneous random walks and subdiffusive transport of cells,
arXiv:1209.2851[cond-mat.stat-mech].


\bibitem{zumofen} G. Zumofen, J. Klafter, and A. Blumen, Chemical Physics \textbf{146},
433 (1990); G. Zumofen and J. Klafter, Phys. Rev. E \textbf{51},
1818, (1995).

\bibitem{prudnikov} A.P. Prudnikov, Yu. A. Brychkov, and O.I.
Marichev, \textit{Integrals and series, Special Functions} (Gordon
and Breach, New York 1986).

\bibitem{SokKlafBlum} I.M. Sokolov, J. Klafter, and A. Blumen,
Phys. Today \textbf{55}(11), 48 (2002).

\bibitem{NIG92} R.R. Nigmatulin,
Theor. Math. Phys. \textbf{90}, 245 (1992).

\bibitem{NIG98} A. Le Mehaute, R. R. Nigmatullin, and L. Nivanen,
\textit{Fleches du Temps et Geometric Fractale} (Hermes, Paris,
1998), Chap. 5.

\bibitem{schulz} J.H.P. Schulz, E. Barkai, and R. Metzler,
Phys. Rev. Lett. \textbf{110}, 020602 (2013).

\bibitem{barkai} E. Barkai, Phys. Rev. Lett. \textbf{90},
104101 (2003).

\bibitem{iom2011} A. Iomin,
Phys. Rev. E \textbf{83}, 052106 (2011).

\bibitem{Cox} D.R. Cox and H.D. Miller, \textit{The Theory of
Stochastic Processes} (Methuen \& CO LTD, London, 1970).

\bibitem{shlezinger1} M.F. Shlesinger, J. Satat. Phys. {\bf 10}, 421
(1974).

\bibitem{blumen} A. Blumen, J. Klafter, and G. Zumofen, in {\it Fractals
in Physics}, eds. L. Pietronero and E. Tosatti, (Noth--Holland,
Amsterdam 1986), p. 399.

\bibitem{bi2011a} E. Baskin and A. Iomin,
Chaos, Solitons \& Fractals \textbf{44}, 335, (2011).

\bibitem{feldman} Yu. Polevaya, I. Ermolina, M. Schlesinger,
B.-Z. Ginzburg,and Yu. Feldman,
Biochimica et Biophysica Acta \textbf{1419}, 257 (1999).

\bibitem{Rubinsky}B. Rubinsky, G. Onik, and P. Mikus,
Technol. Cancer Res. Treat. \textbf{6}(1), 37 (2007).

\bibitem{Shalaev} A.K. Sarychev  and V.M. Shalaev,
\textit{Electrodynamics of metamaterials} (World Scientic,
Singapore, 2007).

\bibitem{Stockman} M.I. Stockman, Physics Today \textbf{64}(2),
39 (2011).

\bibitem{bi2011b} E. Baskin and A. Iomin,
Europhys. Lett. \textbf{96}, 54001 (2011).

\bibitem{Ren}J.R. Liang, X.T. Wang, and W.Y. Qiu,
Chaos, Solitons, \& Fractals \textbf{16}, 107 (2003).

\bibitem{berrypercival} M.V. Berry and I.C. Percival,
Optica Acta \textbf{33}, 577 (1986).

\bibitem{benavraam} D. ben-Avraam and S. Havlin, \textit{Diffusion
and Reactions in Fractals and Disodered Systems} (University
Press, Cambridge, 2000).


\end{thebibliography}
\end{document}